\documentclass[12pt, journal,compsoc]{IEEEtran}
\ifCLASSOPTIONcompsoc
\else
\fi
\ifCLASSINFOpdf
\else
   \usepackage[dvips]{graphicx}
   \DeclareGraphicsExtensions{.eps}
\fi

\usepackage{balance}

\usepackage[ruled]{algorithm2e}

\usepackage[tight,normalsize,sf,SF]{subfigure}

\usepackage{algorithmic}
\begin{document}
%
% paper title
% can use linebreaks \\ within to get better formatting as desired
\title{Energy-aware Fixed-Priority Multi-core Scheduling for Real-time Systems}
%
%
% author names and IEEE memberships
% note positions of commas and nonbreaking spaces ( ~ ) LaTeX will not break
% a structure at a ~ so this keeps an author's name from being broken across
% two lines.
% use \thanks{} to gain access to the first footnote area
% a separate \thanks must be used for each paragraph as LaTeX2e's \thanks
% was not built to handle multiple paragraphs
%
%
%\IEEEcompsocitemizethanks is a special \thanks that produces the bulleted
% lists the Computer Society journals use for "first footnote" author
% affiliations. Use \IEEEcompsocthanksitem which works much like \item
% for each affiliation group. When not in compsoc mode,
% \IEEEcompsocitemizethanks becomes like \thanks and
% \IEEEcompsocthanksitem becomes a line break with idention. This
% facilitates dual compilation, although admittedly the differences in the
% desired content of \author between the different types of papers makes a
% one-size-fits-all approach a daunting prospect. For instance, compsoc
% journal papers have the author affiliations above the "Manuscript
% received ..."  text while in non-compsoc journals this is reversed. Sigh.

\author{Yao~Guo,
        Junyang~Lu% <-this % stops a space
\IEEEcompsocitemizethanks{\IEEEcompsocthanksitem Yao Guo and Junyang Lu are with the Key Laboratory of High-Confidence Software Technologies (Ministry of Education), School of Electronics Engineering and Computer Science, Peking University, Beijing 100871, China.\protect\\

% note need leading \protect in front of \\ to get a newline within \thanks as
% \\ is fragile and will error, could use \hfil\break instead.
E-mail: \{yaoguo, lujunyang\}@pku.edu.cn}% <-this % stops a space

\thanks{This manuscript is an extension of our previous conference paper published at RTCSA 2011~\cite{Lu-RTCSA13}.}}

% note the % following the last \IEEEmembership and also \thanks -
% these prevent an unwanted space from occurring between the last author name
% and the end of the author line. i.e., if you had this:
%
% \author{....lastname \thanks{...} \thanks{...} }
%                     ^------------^------------^----Do not want these spaces!
%
% a space would be appended to the last name and could cause every name on that
% line to be shifted left slightly. This is one of those "LaTeX things". For
% instance, "\textbf{A} \textbf{B}" will typeset as "A B" not "AB". To get
% "AB" then you have to do: "\textbf{A}\textbf{B}"
% \thanks is no different in this regard, so shield the last } of each \thanks
% that ends a line with a % and do not let a space in before the next \thanks.
% Spaces after \IEEEmembership other than the last one are OK (and needed) as
% you are supposed to have spaces between the names. For what it is worth,
% this is a minor point as most people would not even notice if the said evil
% space somehow managed to creep in.

% The paper headers
\markboth{}%
{}
% The only time the second header will appear is for the odd numbered pages
% after the title page when using the twoside option.
%
% *** Note that you probably will NOT want to include the author's ***
% *** name in the headers of peer review papers.                   ***
% You can use \ifCLASSOPTIONpeerreview for conditional compilation here if
% you desire.

% The publisher's ID mark at the bottom of the page is less important with
% Computer Society journal papers as those publications place the marks
% outside of the main text columns and, therefore, unlike regular IEEE
% journals, the available text space is not reduced by their presence.
% If you want to put a publisher's ID mark on the page you can do it like
% this:
%\IEEEpubid{0000--0000/00\$00.00~\copyright~2007 IEEE}
% or like this to get the Computer Society new two part style.
%\IEEEpubid{\makebox[\columnwidth]{\hfill 0000--0000/00/\$00.00~\copyright~2007 IEEE}%
%\hspace{\columnsep}\makebox[\columnwidth]{Published by the IEEE Computer Society\hfill}}
% Remember, if you use this you must call \IEEEpubidadjcol in the second
% column for its text to clear the IEEEpubid mark (Computer Society jorunal
% papers don't need this extra clearance.)

% use for special paper notices
%\IEEEspecialpapernotice{(Invited Paper)}

% for Computer Society papers, we must declare the abstract and index terms
% PRIOR to the title within the \IEEEcompsoctitleabstractindextext IEEEtran
% command as these need to go into the title area created by \maketitle.
\IEEEcompsoctitleabstractindextext{%
\begin{abstract}
%\boldmath
Multi-core processors are becoming more and more popular in embedded and real-time systems. While fixed-priority scheduling with task-splitting in real-time systems are widely applied, current approaches have not taken into consideration energy-aware aspects such as dynamic voltage/frequency scheduling (DVS). In this paper, we propose two strategies to apply dynamic voltage scaling (DVS) to fixed-priority scheduling algorithms with task-splitting for periodic real-time tasks on multi-core processors. The first strategy determines voltage scales for each processor after scheduling (Static DVS), which ensures all tasks meet the timing requirements on synchronization. The second strategy adaptively determines the frequency of each task before scheduling (Adaptive DVS) according to the total utilization of task-set and number of cores available. The combination of frequency pre-allocation and task-splitting makes it possible to maximize energy savings with DVS. Simulation results show that it is possible to achieve significant energy savings with DVS while preserving the schedulability requirements of real-time schedulers for multi-core processors.

%Among all techniques studied, the proposed pre-DVS algorithm has demonstrated better results in both schedulability and energy consumption in comparison to the other approaches.
\end{abstract}

% IEEEtran.cls defaults to using nonbold math in the Abstract.
% This preserves the distinction between vectors and scalars. However,
% if the journal you are submitting to favors bold math in the abstract,
% then you can use LaTeX's standard command \boldmath at the very start
% of the abstract to achieve this. Many IEEE journals frown on math
% in the abstract anyway. In particular, the Computer Society does
% not want either math or citations to appear in the abstract.

% Note that keywords are not normally used for peerreview papers.
\begin{IEEEkeywords}
Real-time systems, multi-core scheduling, dynamic voltage scaling (DVS), energy optimization.
\end{IEEEkeywords}}

% make the title area
\maketitle

% To allow for easy dual compilation without having to reenter the
% abstract/keywords data, the \IEEEcompsoctitleabstractindextext text will
% not be used in maketitle, but will appear (i.e., to be "transported")
% here as \IEEEdisplaynotcompsoctitleabstractindextext when compsoc mode
% is not selected <OR> if conference mode is selected - because compsoc
% conference papers position the abstract like regular (non-compsoc)
% papers do!
\IEEEdisplaynotcompsoctitleabstractindextext
% \IEEEdisplaynotcompsoctitleabstractindextext has no effect when using
% compsoc under a non-conference mode.

% For peer review papers, you can put extra information on the cover
% page as needed:
% \ifCLASSOPTIONpeerreview
% \begin{center} \bfseries EDICS Category: 3-BBND \end{center}
% \fi
%
% For peerreview papers, this IEEEtran command inserts a page break and
% creates the second title. It will be ignored for other modes.
\IEEEpeerreviewmaketitle

\section{Introduction}
% Computer Society journal papers do something a tad strange with the very
% first section heading (almost always called "Introduction"). They place it
% ABOVE the main text! IEEEtran.cls currently does not do this for you.
% However, You can achieve this effect by making LaTeX jump through some
% hoops via something like:
%
%\ifCLASSOPTIONcompsoc
%  \noindent\raisebox{2\baselineskip}[0pt][0pt]%
%  {\parbox{\columnwidth}{\section{Introduction}\label{sec:introduction}%
%  \global\everypar=\everypar}}%
%  \vspace{-1\baselineskip}\vspace{-\parskip}\par
%\else
%  \section{Introduction}\label{sec:introduction}\par
%\fi
%
% Admittedly, this is a hack and may well be fragile, but seems to do the
% trick for me. Note the need to keep any \label that may be used right
% after \section in the above as the hack puts \section within a raised box.

% The very first letter is a 2 line initial drop letter followed
% by the rest of the first word in caps (small caps for compsoc).
%
% form to use if the first word consists of a single letter:
% \IEEEPARstart{A}{demo} file is ....
%
% form to use if you need the single drop letter followed by
% normal text (unknown if ever used by IEEE):
% \IEEEPARstart{A}{}demo file is ....
%
% Some journals put the first two words in caps:
% \IEEEPARstart{T}{his demo} file is ....
%
% Here we have the typical use of a "T" for an initial drop letter
% and "HIS" in caps to complete the first word.
Multi-core processors have been adopted not only in high-performance servers and personal computers, but also for embedded and real-time systems. Many real-time scheduling algorithms for multi-core processors have been proposed in recent years, among which the semi-partitioned fixed-priority multi-core scheduling algorithms~\cite{Guan:2010:FMS:1828428.1829220,Lakshmanan:2009:PFP:1581378.1581523} could achieve higher utilization bound.

On the other hand, energy consumption is critical for many battery-operated embedded and real-time systems. However, although techniques such as dynamic voltage/frequency scaling (DVS)~\cite{WeiserWDS94} have been available in most modern processors, these energy-aware aspects have not been considered in the recently proposed semi-partitioned fixed-priority multi-core scheduling algorithms with high utilization bound.

In order to understand the energy implications of these semi-partitioned fixed-priority multi-core scheduling algorithms, this paper explores the possibility of applying DVS to two recently proposed semi-partitioned fixed-priority multi-core scheduling algorithms: SPA2 \cite{Guan:2010:FMS:1828428.1829220} introduced by Guan \emph{et al.} and PDMS\_HPTS\_DS (PHD) \cite{Lakshmanan:2009:PFP:1581378.1581523} introduced by Lakshmanan \emph{et al}. Among the two techniques, SPA2 could reach a utilization bound of 69.3\% and PHD 65\%, both considerably higher compared to the previous approaches in priority-based multi-core scheduling.

Because neither algorithm has considered energy in their approaches, we introduce two different methods to apply DVS to the above two multi-core scheduling algorithms (SPA2 and PHD):

\begin{itemize}

\item \textbf{Static DVS}: We first develop an extended DVS algorithm based on the traditional DVS algorithm for fixed-priority scheduler \cite{Pillai:2001:RDV:502034.502044}, and apply it after scheduling with task-splitting to evaluate the schedulability and energy savings. To be specific, we set the maximal frequency for certain split sub-tasks, so that all the former sub-tasks could finish before the latter ones release. This ensures that the scheduling will meet the timing requirements on synchronization for split tasks.

\item \textbf{Adaptive DVS}: We then propose a new algorithm, which adaptively determines the frequency of each task before scheduling in order to achieve better performance on both schedulability and energy consumption. Adaptive DVS actually schedules the tasks with prolonged execution time based on DVS,  while ensuring that all the tasks meet the timing requirements after DVS. Specifically, the frequency is determined by the total utilization of task-set and the number of cores available, pursuing the maximal balance of tasks in each processor and therefore getting considerably more energy saving compared to the static DVS approach.

\end{itemize}

In order to evaluate the two approaches mentioned above, we developed a simulator to compare their schedulability and energy consumption with different configurations including different total utilization and different number of processor cores.

In terms of schedulability, PHD performs much better than SPA2 when the utilization is over 70\% (Both algorithms can be one hundred percent schedulable when the utilization is below 65\% due to their utilization bounds). PHD can keep its schedulability at more than 90\% when the utilization reaches 90\%, while SPA2 sharply decreases to zero.

When considering their energy consumption, PHD with static DVS algorithm nearly gets close to the worst case. PHD with adaptive DVS saves most energy because of its equal distribution of tasks to every core available. Overall, the simulation results show that PHD with adaptive DVS algorithm has demonstrated lower energy consumption compared to the other three strategies while maintaining the good schedulability of PHD.

The main contribution of this paper is that we explored the possibility of applying DVS to semi-partitioned fixed-priority multi-core scheduling algorithms. We have presented two different approaches to apply DVS, and simulation results shows that energy-aware scheduling is achievable without affecting the schedulability of two state-of-the-art algorithms. To the best of our knowledge, this is the first work considering energy-efficient multi-core scheduling with task-splitting.

The rest of the paper is organized as follows: Section \ref{prelim} introduces the background information and assumptions, as well as notations used in this paper. The proposed static DVS and adaptive DVS algorithms are presented in Section \ref{post} and \ref{pre} respectively. Section \ref{sim} presents the experimental evaluation of scheduling and DVS schemes. In Section \ref{rel}, we review the related work, which focuses on multi-core scheduling for real-time system and energy-aware approaches. We conclude the paper with Section \ref{con}.

%\hfill mds

%\hfill January 11, 2007

\section{Preliminaries}
\label{prelim}

This section introduces the background information and assumptions, as well as notations used in this paper.

\subsection{Task Model}

We shall use the following notation throughout this paper. We consider a \emph{task-set} $\{T_1, T_2, \cdots, T_N\}$ comprising of $N$ periodic tasks. This task-set is assigned to $M$ processor cores. We use the classical $(C,P,D)$ model to represent the parameters of a task $T$, where $C$ represents the worst-case computation time at maximal frequency $f_{max}$ of each job of $T$, $P$ represents the period of T , and $D$ is defined as the deadline of each job of $T$ relative to job release time. For a task without split, its deadline $D$ equals to its period $P$. When it comes to a subtask of a split task, its deadline $D$ is less than its period $P$, in order to set aside time for other subtasks from the same split task.

Tasks $T_i$ : $(C_i,P_i,D_i)$ are ordered such that $i < j$ implies $D_i < D_j$ . Since our proposed algorithms use deadline-monotonic scheduling as the scheduling algorithm on each processor, we can use the task indices to represent the task priorities, i.e., $i$ has higher priority than $j$ if and only if $i < j$. The utilization of each task $i$ is defined as $U_i = C_i / P_i$. The total utilization $U_{tot}$ is given by $\sum\limits_{i=1}^N U_i$.

During scheduling, some tasks are split and assigned to different processors. We call these tasks \emph{split tasks}, which are split into several \emph{subtasks}. For a split task $T_i$, $T_i^k$ denotes the $k$th subtask of $T_i$. We define the last subtask of $T_i$ its \emph{tail subtask}, and other subtasks are called \emph{body subtasks}.

The subtasks of a split task need to be synchronized to execute correctly, which means that $T_i^{k+1}$ cannot start execution until $T_i^k$ is finished. Therefore, the time for a subtask $T_i^{k}$ to execute is shorter than its period, in order to share time with other subtasks from the same split task. For a subtask $T_i^k$ split from task $T_i$, its deadline $D_i^k$ satisfies the equation $D_i^k = P_i - \sum\limits_{j = 1}^{k - 1}R_i^j$, where $R_i^k$ means the actual time span of the subtask from its release to completion.

\subsection{Energy Model}
For processors based on the CMOS technology, the power consumption of each core $P(f)$ consists of two parts: dynamic power dissipation $P_d(f)$ and static power dissipation $P_{ind}$. Dynamic power dissipation $P_d(f)$ is given by: $P_d(f) = C_{ef} V_{dd}^2 f$, where $V_{dd}$ is the supply voltage, $C_{ef}$ is the effective switching capacitance, and $f$ is the processor clock frequency. For simplicity, processor frequency can be considered roughly linearly to the supply voltage: $f = k \frac{(V_{dd}-V_t)^2}{V_{dd}}$, where $k$ is a constant and $V_t$ is the threshold voltage \cite{Chandrakasan95lowpower}. Thus, $P_d(f)$ is almost cubically related to $f$: $P_d(f) \approx C_{ef} \frac{f^3}{k^2}$. Static power dissipation $P_{ind}$ is dominated by the leakage current, and can be regarded as a constant.
%Since the time needed for a specific task is $time = \frac{C}{f}$, where $C$ is the worst-case computation time to execute the task, the energy consumption of the task, $E$, is $E = P_d time \approx  C_{ef} \frac{f^2}{k^2}$.

For simplicity of presentation, we assume the power dissipation function as below:

\begin{equation}
\label{Pafb}
P(f) = \alpha {f^3} + \beta
\end{equation}

where $\alpha$ and $\beta$ are non-negative constants.

The total power consumption of a multi-core processor is simply the sum of the power dissipated in each core: $P_{tot} = \sum\limits_{i=1}^M P_i$. For real-time systems, because we can assume that the processors are always running, the energy consumption is actually proportional to power dissipation in all cases.

\begin{table}[!t]
\renewcommand{\arraystretch}{1.3}
\caption{Frequency/voltage settings of the XScale processor.}
\label{fp}
\centering
\begin{tabular}{|c||c|c|c|c|c|}
\hline
Frequency(MHz) & 150 & 400 & 600 & 800 & 1000\\
\hline
Voltage(V) & 0.75 & 1.0 & 1.3 & 1.6 & 1.8\\
\hline
\end{tabular}
\end{table}

Based on the frequency/power settings of Intel processor XScale\cite{Xu:2004:PPE:1017753.1017767} in Table \ref{fp}, we estimate the parameters of power consumption model in equation (\ref{Pafb}) as below:

\begin{equation}
\label{Pafbp}
P(f) = 1.52 {f^3} + 0.08
\end{equation}

To eliminate the effect of leakage power consumption, we introduce a critical speed \cite{Chen:2007:PDP:1326073.1326132} as the minimal frequency that a processor core can execute at. Based on the model in equation (\ref{Pafbp}), we can calculate that the critical speed, as well as the minimal frequency $f_{min}$, is 0.297GHz.

%When decreasing processor speed, we can also reduce the supply voltage. This reduces processor power cubically and energy quadratically at the expense of linearly increasing the task's latency.

To model different processor frequencies, we have simulated both continuously changed frequencies among $[f_{min}, f_{max}]$ (the ideal case) and also a discrete frequency level model based on an actual processor model (i.e., Intel XScale).

\subsection{Scheduling Algorithms Studied}

As mentioned earlier, this paper studies two existing multi-core scheduling algorithms, on which we will give a brief description respectively.

\subsubsection{SPA2}

The algorithm SPA2 \cite{Guan:2010:FMS:1828428.1829220} could reach Liu \& Layland's Utilization Bound \cite{Liu:1973:SAM:321738.321743} for task sets without any constraints. It assigns tasks in increasing order of priority, and each time selects the processor with the minimal utilization to assign. In order to ensure the schedulability of task-set with heavy tasks, it first assigns heavy tasks that satisfy a particular condition so that the heavy tasks will not be split and the execution sequence will be guaranteed.

\subsubsection{PDMS\_HPTS\_DS}
The algorithm PDMS\_HPTS\_DS (PHD) \cite{Lakshmanan:2009:PFP:1581378.1581523} has a slightly lower utilization bound (65\%) than the algorithm SPA2 (69.3\%). Based on our simulation results, PHD could achieve an average schedulable utilization of 88\%, which is considerably higher than SPA2.
%From the practical perspective, the schedulability of algorithm PHD is better than SPA2.

PHD assigns the tasks in decreasing order of utilization, and assigns tasks to the next processor only when the previous processor could no longer hold more tasks. In order to ensure the sequence rules of split task, the algorithm only split the task with highest priority of each processor so that the sub-tasks of the split task could be completed in sequence.

\section{The Static DVS Approach}
\label{post}

\begin{figure}[t]
\centering
\includegraphics[width=3.5in]{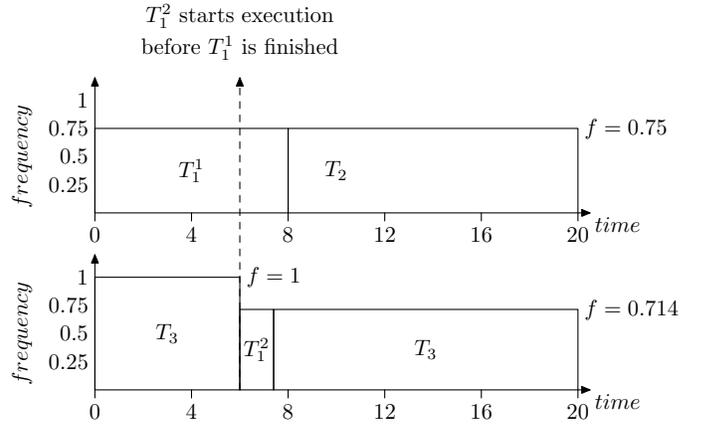}
\caption{Synchronous Violation with Traditional DVS}
\label{SVTD}
\end{figure}

Pillai and Shin have introduced a DVS algorithm for fixed-priority schedulers \cite{Pillai:2001:RDV:502034.502044}, achieving energy savings by reducing the operating frequency and voltage when remaining tasks need less than the remaining time before the next deadline. However, for scheduling with task-splitting, one cannot reduce the frequency freely, because the synchronous requirements of split-tasks may be violated when postponing the execution time of each subtask, as the example shown in Fig. \ref{SVTD}.

In order to achieve energy savings for scheduling with task-splitting, we develop a new static DVS algorithm based on the previous work, which did not take task-splitting into consideration. We reselect the frequency when any of the tasks is released or finishes.
%according to the available time until the earlier one between next deadline and next release time, and the remaining total cycles current unfinished tasks still need.
Before selecting frequencies, we first examine whether the task/subtask is a body subtask. If it is a body subtask, we   it under maximal frequency and reselect the frequency when this subtask finishes, so that the synchronous requirement is always satisfied.

The detailed description of the static DVS algorithm for scheduling with task-splitting is shown in Algorithm \ref{ag-post}. In function $available\_time\_until\_next\_time\_line()$, the time line represents either the release time or deadline. So the function returns the smaller one between the available time until the next deadline and the available time until the next release time.

\begin{algorithm}
\caption{Static DVS for Scheduling with Task-Splitting}
\label{ag-post}
\begin{algorithmic}[1]
\STATE \textbf{select\_frequency():}
\STATE $s_m:=$available\_time\_until\_next\_time\_line()
\STATE $f:=f_{max}*\sum d_i/s_m$
\vspace{0.05in}
\STATE \textbf{upon task\_release($T_i$):}
\STATE $C\_left_i:=C_i$
\STATE $s_m:=$available\_time\_until\_next\_time\_line()
\STATE allocate\_cycles($s_m$)
\IF {$T_i$ is body subtask}
\STATE $f:=f_{max}$
\ELSE
\STATE select\_frequency()
\ENDIF
\vspace{0.05in}
\STATE \textbf{task\_completion($T_i$):}
\STATE $C\_left_i:=0$
\STATE $d_i:=0$
\STATE select\_frequency()
\vspace{0.05in}
\STATE \textbf{during task\_execution($T_i$):}
\STATE decrement $C\_left_i$ and $d_i$
\vspace{0.05in}
\STATE \textbf{allocate\_cycles(k):}
\FOR{$i:=1$ to $N$}
\IF {$C\_left_i<k$}
\STATE $d_i:=C\_left_i$
\STATE $k:=k-C\_left_i$
\ELSE
\STATE $d_i:=k$
\STATE $k:=0$
\ENDIF
\ENDFOR
\end{algorithmic}
\end{algorithm}

We can prove that the above static DVS algorithm will not violate any of the timing requirements. The timing requirements can be separated as two statements as below.

\begin{enumerate}
\item If a task set is schedulable under a scheduling algorithm without DVS, it could still be ensured that all the tasks finish before their deadlines after applying the static DVS algorithm.
\item The split task could satisfy the synchronous requirements. That is, it is guaranteed that the next sub-task will be released after the previous one finishes.
\end{enumerate}

For the first statement, we take into consideration the earliest time from now that may break a deadline. This earliest time may be a task's deadline or a task's release time which gains the remaining cycles to execute and therefore increasing the chance to violate a later deadline. While taking this earliest time as a bound while selecting frequency, it is guaranteed that all the deadlines can be met.

For the second statement, we notice that the two algorithms \cite{Guan:2010:FMS:1828428.1829220} \cite{Lakshmanan:2009:PFP:1581378.1581523} we studied both satisfy the following property: the body subtask always has the highest priority in its core, so that they can complete for the next subtask as soon as possible. The tail task only needs to meet its deadline. So all we need to do is allowing these body subtasks execute with no latency compared to the original schedule without DVS. Executing the body subtask at the maximal frequency can guarantee this.

\begin{figure}[!t]
\centering
\subfigure[Energy for SPA2]{
\includegraphics[width=1.7in]{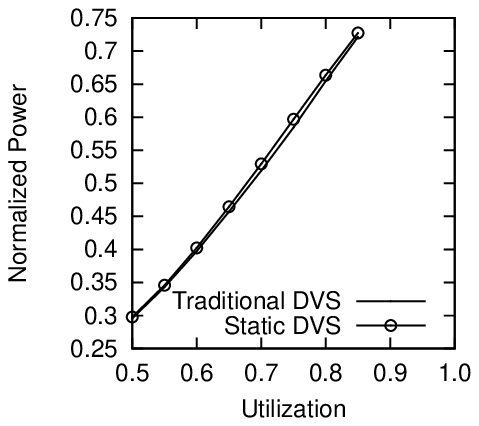}}
\hspace{-0.2in}
\subfigure[Energy for PHD]{
\includegraphics[width=1.7in]{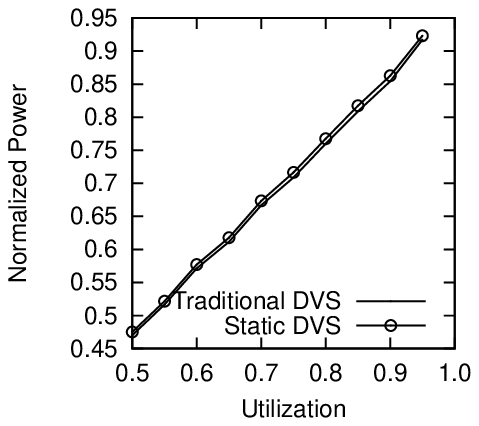}}
\caption{Energy consumption between traditional DVS and our DVS}
\label{TDOD}
\end{figure}

On the other hand, the energy savings suffer very little through our trade-off for split tasks. We performed a simple simulation for scheduling with eight cores, comparing the energy consumption between traditional DVS and the proposed static DVS algorithm for scheduling with task-splitting. The simulation results are shown in Fig. \ref{TDOD}. It shows clearly that there are only slight differences on energy consumption between traditional DVS and the static DVS algorithm. Actually, the energy consumption of static DVS is 1.2\% more than that of the traditional DVS approach on average. We believe that such a small cost is acceptable in order to meet the synchronization requirements for split tasks.

The static DVS approach will be evaluated in detail later in Section \ref{sim}.

\section{The Adaptive DVS Approach}
\label{pre}

To get the best results of DVS, we attempt explore better algorithms to achieve more energy savings compared to the above static DVS algorithm.

In this section, we first consider the potential of energy optimization, and then propose a new DVS algorithm which adaptively determines the frequency of each task before scheduling (Adaptive DVS) that can save more energy compared to the static DVS approach.

\subsection{Energy Optimization}

Given a task-set with periodic real-time tasks and a processor with M cores, we need to find a schedulable task-to-core assignment that minimizes the energy consumption under DVS. Thus, two conditions \cite{Aydin:2003:EPM:838237.838347} must be satisfied:

\begin{enumerate}
\item The assignment must evenly divide the total load $U_{tot}$ among all the cores.
\item In each core with total utilization $S$, the frequency $f$ must be constant and equal to $S$.
\end{enumerate}

That is, each processor must manage to run under constant frequency $f$ that satisfies $f/f_{max}=U_{tot}/M$, where $f_{max}$ is the maximal frequency a processor's multiple supply voltages could provide. We call this frequency the ``ideal frequency'':
\begin{center}
$f_{ideal} = f_{max} U_{tot}/M$\\
\end{center}

With frequency pre-allocation and task-splitting, it is possible to get very close to the minimal energy, as long as the assignment is schedulable.

\subsection{Motivating Example}

Consider three tasks shown in table \ref{edata} to be executed on a 2-core processor. We try to assign the task-set with different algorithms and compute the energy consumption on each processor.

\begin{table}[!t]
\caption{An example task-set}\label{edata}
\centering
\begin{tabular}{|c||c|c|c|}

\hline
& $C_i$ & $P_i$ & $U_i$\\
\hline
$T_1$ & 3 & 5 & 0.6\\
\hline
$T_2$ & 3 & 5 & 0.6\\
\hline
$T_3$ & 4 & 10 & 0.4\\
\hline
\end{tabular}
\end{table}

\subsubsection{SPA2}

According to the algorithm description from Guan \emph{et al.} \cite{Guan:2010:FMS:1828428.1829220}, the task-set is not schedulable under SPA2, because the average utilization for each processor exceeds Liu \& Layland's Utilization Bound \cite{Liu:1973:SAM:321738.321743}:

\begin{center}
$(0.6 + 0.6 + 0.4)/2 > UtilizationBound(3) = 0.78$.
\end{center}

\subsubsection{PHD with Static DVS}

With PHD, the task assignment is shown in Fig. \ref{A1}: The first processor is fully used, while only 60\% of the second processor is used. With static DVS, based on the model in equation (\ref{Pafbp}), the energy consumption in 10 seconds of the first core is $E_1 = \sum t P(f) = 10 P(1) = 10 \alpha + 10 \beta = 16$. The energy consumption in 10 seconds of the second core is $E_2 = \sum t P(f) = 3 P(1) + 4 P(0.5) + 3 P(0.333) = 3.61 \alpha + 10 \beta = 6.2872$. Thus, the total energy is $E_{static} = E_1 + E_2 = 22.2872$.

\subsubsection{PHD with Adaptive DVS}

\begin{figure}[!t]
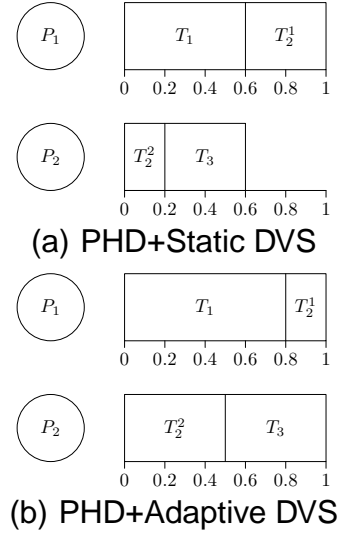

\centering
\subfigure[PHD+Static DVS]{
\includegraphics[width=1.65in]{a.eps}
\label{A1}}
\hspace{-0.1in}
\subfigure[PHD+Adaptive DVS]{
\includegraphics[width=1.65in]{b.eps}
\label{A2}}
\caption{Assignment for task-set in TABLE \ref{edata}}
\end{figure}

If we first decide the frequency $f$ to be $f = U_{tot} / M = 0.8$, the utilization for each task could be considered as (0.75, 0.75, 0.5), because of the postponement of the execution. Then, we adjust such data into the PHD algorithm. The task assignment is shown in Fig. \ref{A2}: Both cores are fully used, and each core's energy consumption in 10 seconds is $E_1 = E_2 = \sum t P(f) = 10 P(0.8) = 5.12 \alpha + 10 \beta = 8.5824$. Thus, the total energy is $E_{adaptive} = E_1 + E_2 = 17.1648$, which is obviously much less than the previous result of $22.2872$. Actually, in this case, the strategy that deciding the frequency before scheduling has satisfied the two conditions for energy minimization mentioned above.
%That is, 1.024 is the minimal total energy for this task-set under $2$ cores.

\vspace{0.1in}

In the example above, we can see that adaptively deciding the frequency before scheduling (Adaptive DVS) consumes much less energy than directly assigning tasks in a depth-first way (as PHD with static DVS does). At the same time, PHD with adaptive DVS manages to assign task-sets with larger total utilization than SPA2.

\begin{figure}[!t]
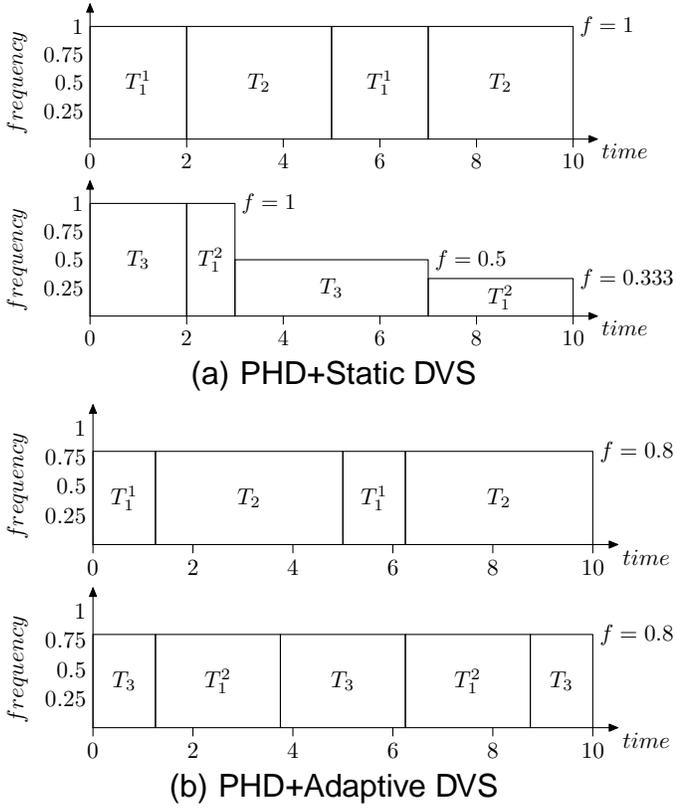

\centering
\subfigure[PHD+Static DVS]{
\includegraphics[width=3.5in]{ae.eps}}
\subfigure[PHD+Adaptive DVS]{
\includegraphics[width=3.5in]{be.eps}}
\caption{Execution for task-set in TABLE \ref{edata}}
\end{figure}

\subsection{Algorithm Description}

Even though the two conditions for energy minimization are usually hard to satisfy, we could still try to achieve a result as close as possible. Below are the detailed steps of the proposed adaptive DVS algorithm.

\textbf{Step 1.} We try to set all the tasks' execution frequency to the same ideal level: $f = f_{ideal} = f_{max} U_{tot}/M$.

\textbf{Step 2.} Because some tasks may not execute in such low frequency due to their relatively high utilization, which might be greater than $f/f_{max}$, we deal with the tasks in two different ways: if a task $T_i$ satisfies $U_i > f/f_{max}$, we set its frequency $f_i$ as $f_{max} U_i$. Otherwise, we set its frequency $f_i$ as $f$.

\textbf{Step 3.} After changing the frequency of each task, we need to extend their execution time accordingly. For convenience in the next step, we regard the extended execution time as new execution time, and store the original one. Thus, for each task, $C_i = Old\_C_i$.

\textbf{Step 4.} We try the new task-set with the scheduling algorithm (PHD for example). If it is schedulable, we can decide the minimal $f$ that is schedulable for certain task-set and processors. Otherwise, we need to gradually increase the frequency $f$ and repeat Step 2, 3, 4 until it is schedulable.

As a result, we could assign the tasks to processors as balanced as possible and achieve significant energy savings with the adaptive DVS algorithm. Experiments in the next section will show that such an algorithm could maintain the good schedulability of PHD while consuming much less energy.

The detailed description of the adaptive DVS algorithm for scheduling with task-splitting is shown in Algorithm \ref{ag-pre}.

\begin{algorithm}
\caption{Adaptive DVS for Scheduling with Task-Splitting}
\label{ag-pre}
\begin{algorithmic}[1]
\FOR{each $f$ available among the range of [$f_{max} U_{tot}/M$, $f_{max}$]}
\FOR{each $i\in [1,N]$}
\STATE $Old\_C_i:=C_i$
\ENDFOR
\FOR{each $i\in [1,N]$}
\IF{$U_i<=f/f_{max}$}
\STATE $f_i:=f$
\STATE $C_i:=C_i f_{max}/f$
\ELSE
\STATE $f_i:=f_{max} U_i$
\STATE $C_i:=P_i$
\ENDIF
\ENDFOR
\IF{$Schedulable() = TRUE$}
\STATE Done
\ELSE
\FOR{each $i\in [1,N]$}
\STATE $C_i:=Old\_C_i$
\ENDFOR
\ENDIF
\ENDFOR
\end{algorithmic}
\end{algorithm}

\section{Simulation}
\label{sim}

We have developed a simulator to evaluate the schedulability and energy savings from dynamic voltage scaling in a multi-core real-time system for both static DVS and adaptive DVS approaches on the two scheduling algorithms SPA2 and PHD.

\subsection{Simulation Methodology}

\begin{figure}[!t]
\centering
\includegraphics[width=2.5in]{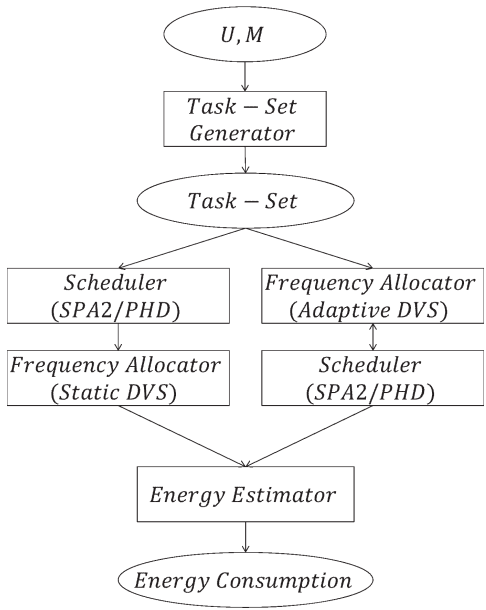}
\caption{The working flow of our simulator.}
\label{so}
\end{figure}

We developed a simulator for the operation of hardware capable of voltage and frequency scaling with real-time scheduling. This simulator takes average utilization per core and number of cores as input, and calculates the schedulability and energy consumption for each of the algorithms we have studied: SPA2 with static DVS, PHD with static DVS, SPA2 with adaptive DVS and PHD with adaptive DVS. Figure \ref{so} illustrates the outline of the simulator, which consists of four components - task set generator, scheduler of different algorithms, frequency allocator under different time and energy estimator.

The real-time tasks are specified using pairs of $(C, P)$, indicating their worst-case computation time and period. The task-sets are generated as follows. Each task has an equal probability of having a period among [10, 20, 30, 40, 50, 60, 70, 80, 90, 100, 200, 300, 400, 500, 600, 700, 800, 900, 1000]. This simulates the varied mix of short and long period tasks commonly found in real-time systems. The worst-case computation time is uniformly distributed over [0, P]. Finally, the last task computation requirements are scaled by a constant chosen such that the sum of the utilizations of the tasks in the task-set reaches a desired value. We simulate each task on platforms with 2, 4, 8 or 16 cores. For a fixed number of cores $M$, we varied average utilization per core among $[0.1, 0.2, \cdots, 1.0]$.

\subsection{Energy Calculation}

To achieve an ideal effect from DVS, we first assume that multiple supply voltages is continuous. That is, the processor could execute in any frequency in range [$f_{min}$, $f_{max}$], where $f_{min}$ is the critical speed.

Then, for more practical purpose, we simulate the approaches under discrete voltages and frequency. To model a real processor, we use the frequency/power settings of Intel XScale \cite{Xu:2004:PPE:1017753.1017767}(as shown in Table \ref{fp}) for discrete frequencies.

Only energy consumed by the processor is computed, and variations due to different types of instructions executed are not taken into account. With this simplification, the task execution model can be reduced to counting cycles of execution, and execution traces are not needed. In particular, this does not consider preemption and task switching overheads, or the time required to switch operating frequency or voltages. There is no loss of generality from these simplifications. The preemption and task switch overheads are the same with static DVS or adaptive DVS, so they have no (or very little) effect on relative energy consumption numbers.

%The independent variable $\sum U_i/M$ ranges from 0.1 to 1, and $M$ ranges among [2, 4, 8, 16].

\subsection{Simulation Results}

We performed simulations for the four approaches, which
include the two static DVS approaches (PHD+Static DVS and SPA2+Static DVS),
and the two adaptive DVS approaches (PHD+Adaptive DVS and SPA2+Adaptive DVS).
Each utilization and core number level in the
results corresponds to simulations of 10,000 task-sets.
The results shown are average numbers from 10,000 simulations.

\begin{figure*}[!t]
\centering
\subfigure[2 Cores]{
\includegraphics[width=1.80in]{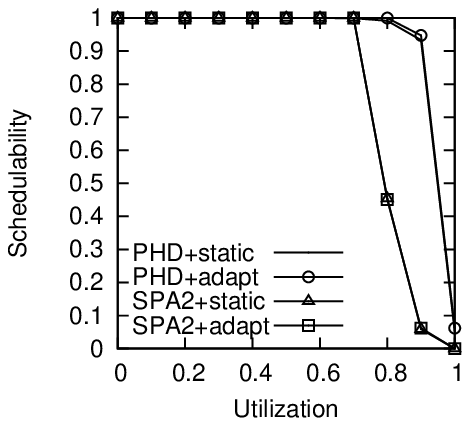}}
\hspace{-0.2in}
\subfigure[4 Cores]{
\includegraphics[width=1.80in]{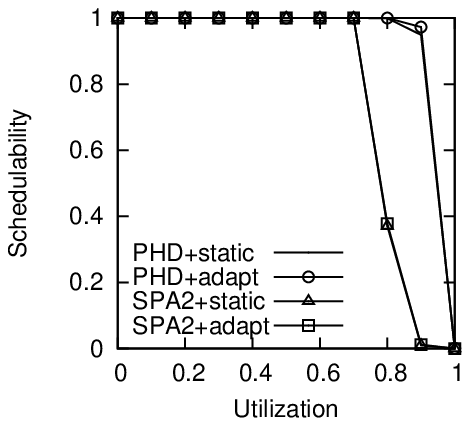}}
\hspace{-0.2in}
\subfigure[8 Cores]{
\includegraphics[width=1.80in]{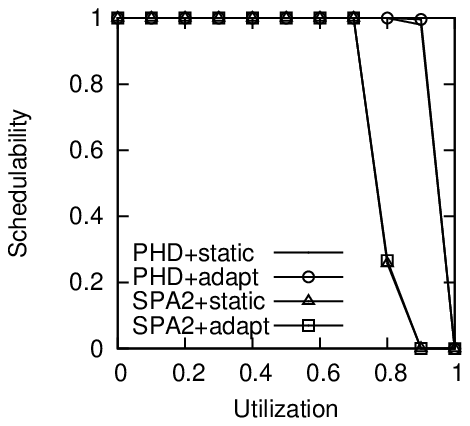}}
\hspace{-0.2in}
\subfigure[16 Cores]{
\includegraphics[width=1.80in]{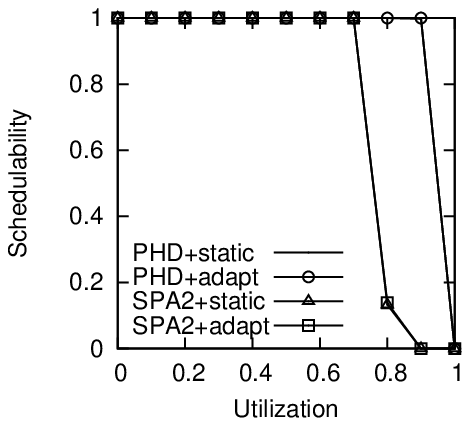}}
\caption{Schedulability simulation results (under continuous voltages/frequencies).}
\label{s}
\end{figure*}

\begin{figure*}[!t]
\centering
\subfigure[2 Cores]{
\includegraphics[width=1.80in]{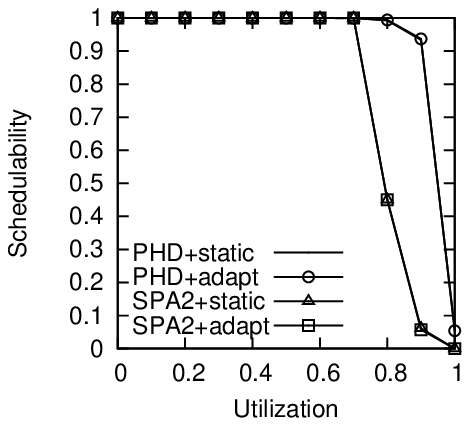}}
\hspace{-0.2in}
\subfigure[4 Cores]{
\includegraphics[width=1.80in]{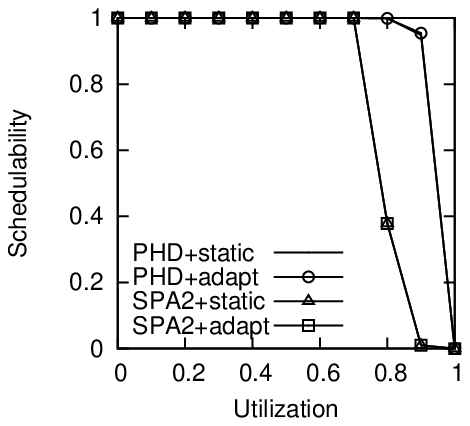}}
\hspace{-0.2in}
\subfigure[8 Cores]{
\includegraphics[width=1.80in]{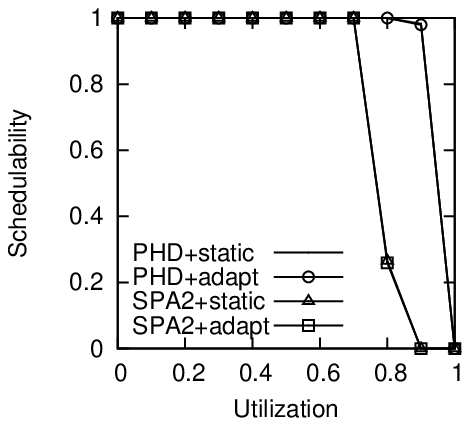}}
\hspace{-0.2in}
\subfigure[16 Cores]{
\includegraphics[width=1.80in]{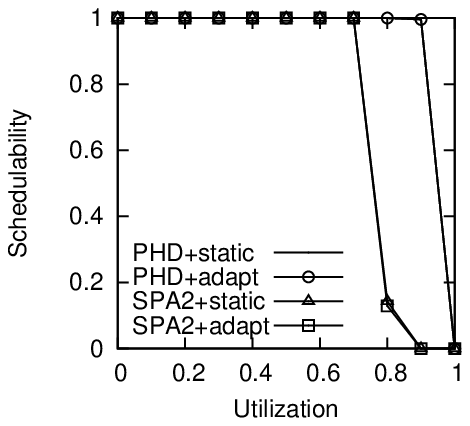}}
\caption{Schedulability simulation results (under discrete voltages/frequencies).}
\label{ds}
\end{figure*}

\subsubsection{Schedulability}

Fig. \ref{s} shows simulation results of schedulability for all the
approaches under continuous frequencies/voltages.
Schedulability stands for the possibility that a task-set with specified total utilization is schedulable to a specified number of processors under an algorithm.

From Fig. \ref{s}, we can see clearly that all the algorithms have a schedulability of 100\% when the average utilization is under 70\%, which corresponds to the utilization bound given by the previous work -- 65\% for PHD and 69.3\% for SPA2. However, the two scheduling algorithms show a remarkable difference when the average utilization exceeds 70\%. SPA2 does not perform so well as its utilization bound, due to its severe restrictions set by Liu \& Layland's Utilization Bound \cite{Liu:1973:SAM:321738.321743} during scheduling. On the contrary, PHD has a much better schedulability at higher utilization because it fills every processor as full as possible without unnecessary restrictions. Of course, it is difficult for either algorithm to schedule most of task-sets when the total utilization equals to the number of processors. So the schedulability of almost all algorithms decreases to zero when the average utilization reaches 100\%.

In addition, we realize that the time when we perform DVS (either static DVS or adaptive DVS) does not make much difference on schedulability. After all, adaptive DVS will try all the frequencies until it is schedulable or the frequency becomes maximal as static DVS does. So any task-set that is schedulable with static DVS algorithms could be schedulable with adaptive DVS algorithms as well. On the other hand, adaptive DVS algorithms try to reduce the frequency by prolonging the time a task executes, so it is hard to find a task-set schedulable with adaptive DVS while unschedulable with static DVS, although it does exist in rare instances. As a whole, the schedulability of static DVS and adaptive DVS are at approximately the same level, as shown in the simulation results.

To model a real processor, we also simulate the algorithms with discrete voltages and frequencies as the XScale processor mentioned above. The results of schedulability are shown in Fig. \ref{ds}. The results show that the schedulability under discrete voltages are almost the same as the previous results under continuous voltages. This is because the differences of frequencies are not significant enough to affect the schedulability statistics.

\begin{figure*}[!t]
\centering
\subfigure[2 Cores]{
\includegraphics[width=1.80in]{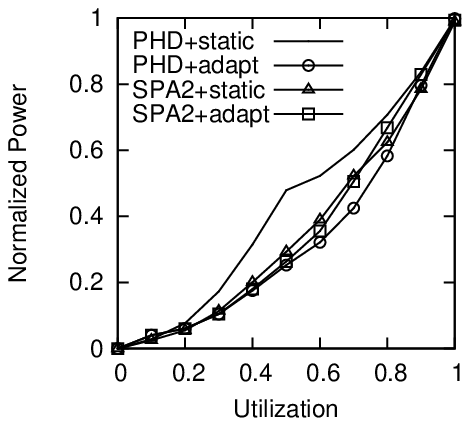}}
\hspace{-0.2in}
\subfigure[4 Cores]{
\includegraphics[width=1.80in]{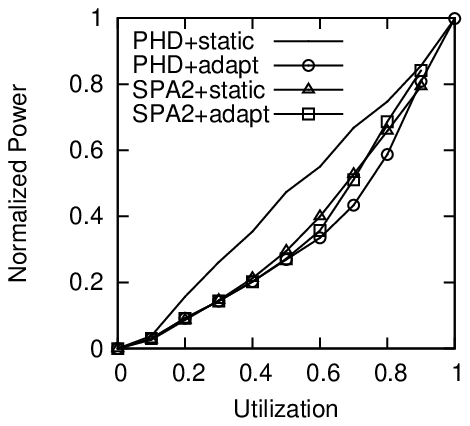}}
\hspace{-0.2in}
\subfigure[8 Cores]{
\includegraphics[width=1.80in]{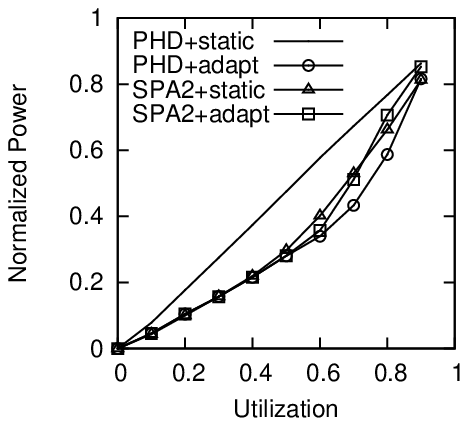}}
\hspace{-0.2in}
\subfigure[16 Cores]{
\includegraphics[width=1.80in]{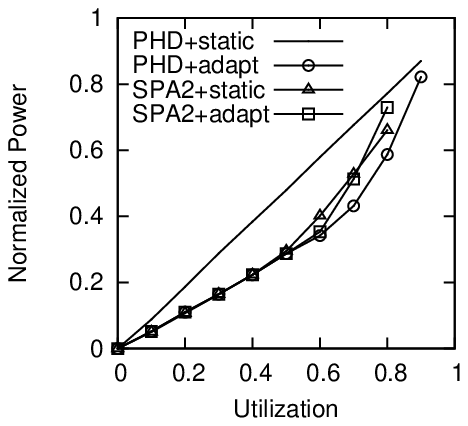}}
\caption{Energy simulation results (under continuous voltages/frequencies).}
\label{p}
\end{figure*}

\begin{figure*}[!t]
\centering
\subfigure[2 Cores]{
\includegraphics[width=1.80in]{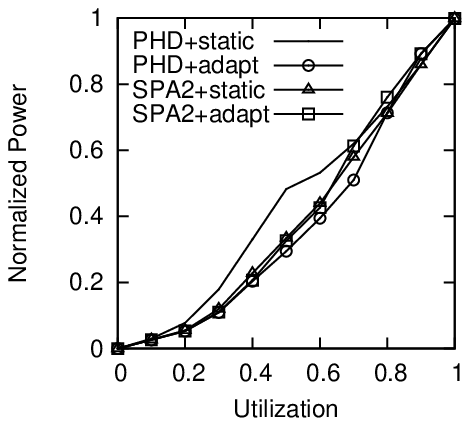}}
\hspace{-0.2in}
\subfigure[4 Cores]{
\includegraphics[width=1.80in]{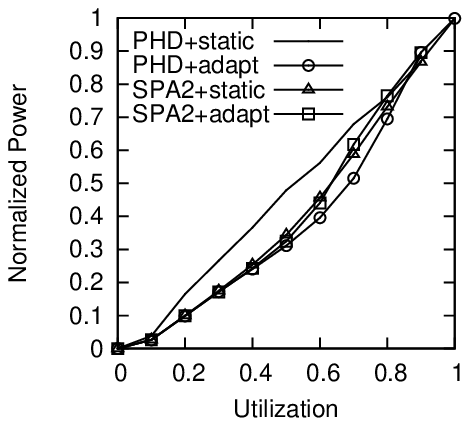}}
\hspace{-0.2in}
\subfigure[8 Cores]{
\includegraphics[width=1.80in]{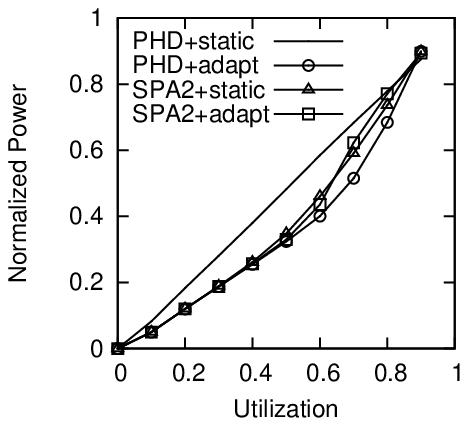}}
\hspace{-0.2in}
\subfigure[16 Cores]{
\includegraphics[width=1.80in]{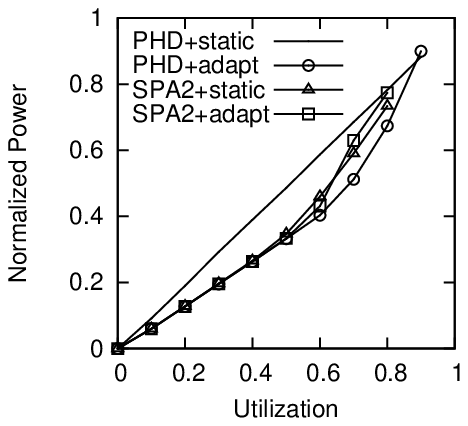}}
\caption{Energy simulation results (under discrete voltages/frequencies).}
\label{dp}
\end{figure*}

\subsubsection{Energy}

Fig. \ref{p} shows the energy numbers for all the approaches.
The energy numbers in our results are normalized to the energy consumed under a processor's maximal frequency.

From Fig. \ref{p}, we notice that when the average utilization is close to 0, the energy numbers of all the algorithms reduce to 0 as expected, because there are few tasks to execute and DVS just reduces the frequency to 0 to save energy (of course this is the ideal case). When the average utilization comes to 100\%, the normalized energy numbers of all the algorithms reach 1, because all the cores have to be full and keep at maximal frequency as long as the task-set is schedulable.

When the utilization gets close to about 50\%, PHD with static DVS gradually shows much higher energy consumption compared to the other three techniques. This is because PHD greedily assigns tasks to as few processors as possible and just keeps other cores idle, while DVS could not take effect when a core is completely full or empty. As a result, the normalized energy of PHD with static DVS always approximately equals to the average utilization per core, showing an almost linear relationship.

%Specifically, the energy of PHD with post-DVS comes to a peak point when the total utilization is an integer, which means the tasks could exactly keep some processors completely full and the others completely empty. At other times, there is always one processor half empty and half full, giving fewer chances for DVS to take effects. However, this effect on a single processor becomes weaker and weaker when the number of total processors increases.

As to the other three algorithms, they have different ways to schedule the tasks evenly to all the processors, and therefore achieve a considerable energy saving. The SPA2 algorithms assign tasks in a width-first way. Adaptive DVS algorithms increase all the tasks utilization as high as possible, so that they have to fill all the cores.

In all the cases, we found that PHD with adaptive DVS achieves the most energy savings, while PHD with static DVS achieves the least energy saving. For example, for the 8-core case with 70\% utlization, PHD with adaptive DVS reduces energy by 56.7\%, compared to only 32.7\% savings for PHD with static DVS. The benefits of the other two techniques (SPA2 with static DVS and adaptive DVS) are roughly 50\%.
The explanation to this is that pre-allocation of frequencies and a depth-first way assigning with task-splitting produce the most balancing scheduling, thus taking most advantages of all the cores.

In a more practical situation, we simulate the algorithms with discrete voltages/frequencies as the XScale processor mentioned above. The results are shown in Fig. \ref{dp}. The energy savings become smaller for all the algorithms compared to the continuous voltage/frequency case. It is because that the frequencies for a processor to choose is much more restricted. Among the four algorithms, the energy savings of PHD with static DVS decreases the least, because it has been to the worst point under continuous frequencies and could not get worse. PHD with adaptive DVS remains to be the best approaches, although the gap between the best and worst technique also shrinks.

\vspace{0.1in}

From the simulation results, we can see that it is practical to apply energy-saving techniques such as DVS to multi-core scheduling algorithms with task-splitting. Although all the four approaches we have studied could save considerable energy consumption with DVS, the PHD scheduling algorithm with adaptive DVS shows both excellent schedulability and energy savings among all the approaches.

\section{Related Work}
\label{rel}
Multi-core scheduling schemes for real-time system can be classified into global and partitioned approaches. In global scheduling, all tasks are put in a global queue and each processor selects from the queue the task with the highest priority for execution. In partitioned scheduling, each task is assigned to a specific processor and each processor fetches tasks for execution from its own queue. It has been shown that each of these categories has its own advantages and disadvantages \cite{Lauzac98comparisonof}. Global scheduling schemes can better utilize the available processors, as illustrated by PFair \cite{Baruah:1998:PSG:626526.627187} and LLREF \cite{Cho:2006:ORS:1193218.1194408}. These schemes appear to be best-suited for applications with small working set sizes. On the other hand, partitioned approaches are severely limited by the low utilization bounds associated with bin-packing problems. The advantage of these schemes is their stronger processor affinity, and hence they provide better average response times for tasks with larger working set sizes.

Global scheduling schemes based on rate-monotonic scheduling (RMS) and earliest deadline first (EDF) are known to suffer from the so-called Dhall effect. When heavyweight (high-utilization) tasks are mixed with lightweight (low-utilization) tasks, conventional real-time scheduling schemes can yield arbitrarily low utilization bounds on multiprocessors. By dividing the task-set into heavy-weight and lightweight tasks, the RM-US \cite{Andersson01static-priorityscheduling} algorithm achieves a utilization bound of 33\% for fixed-priority global scheduling. These results have been improved with a higher bound of 37.5\% \cite{Lundberg:2002:AFG:827265.828503}. The global EDF scheduling schemes have been shown to possess a higher utilization bound of 50\% \cite{Baker:2005:AES:1070609.1070737}. PFair scheduling algorithms based on the notion of proportionate progress \cite{Baruah:1993:PPN:167088.167194} can achieve the optimal utilization bound of 100\%. Despite the superior performance of global schemes, significant research has also been devoted to partitioned schemes due to their appeal for a significant class of applications, and their scalability to massive multi-cores, while exploiting cache affinity.

Partitioned multiprocessor scheduling techniques have largely been restricted by the underlying bin-packing problem. The utilization bound of strictly partitioned scheduling schemes is known to be 50\%. This optimal bound has been achieved for both fixed-priority algorithms \cite{10.1109/EMRTS.2003.1212725} and dynamic priority algorithms based on EDF \cite{Lopez:2004:UBE:1008193.1008208}. Most modern multi-core processors provide some level of data sharing through shared levels of the memory hierarchy. Therefore, it could be useful to split a bounded number of tasks across processing cores to achieve higher system utilization \cite{NizR06}. Partitioned dynamic-priority scheduling schemes with task splitting have been explored in this context \cite{Andersson:2006:MSF:1157741.1158329} \cite{Kato:2007:RST:1306877.1307300}. Partitioned fix-priority scheduling schemes with task-splitting are also explored recently \cite{Lakshmanan:2009:PFP:1581378.1581523} \cite{Guan:2010:FMS:1828428.1829220}. Lakshmanan \emph{et al.}. \cite{Lakshmanan:2009:PFP:1581378.1581523} showed that the cache overheads due to task-splitting can be expected to be negligible on multi-core platforms. However, few works have considered the problem of energy consumption while scheduling with task-splitting.

On the other hand, there are also many works on energy-aware scheduling for real-time system \cite{Aydin:2003:EPM:838237.838347,Mei-2013,Fan-SAC13,Wu-HPCC13}. It is proved that DVS can achieve significant energy savings. But none of the energy-aware techniques have considered the task-splitting strategies, which provides better utilization on the available processors. In this paper, we focus on the energy aspects and explore energy-aware partitioned fix-priority scheduling schemes with task-splitting. Our previous work~\cite{Lu-RTCSA13} has shown preliminary and promising results on this topic.

\section{Conclusion}
\label{con}

In this paper, we have explored the possibility of combining dynamic voltage (frequency) scheduling with semi-partitioned fixed-priority multi-core scheduling with task-splitting for real-time systems. We proposed two different techniques to apply the DVS algorithm to multi-core scheduling approaches with task-splitting features. The techniques proposed include performing DVS after scheduling (Static DVS) and performing DVS before scheduling (Adaptive DVS).

We simulated the proposed techniques under different processor setups.
Simulation results show that it is possible to achieve significant energy savings with DVS while preserving the schedulability requirements of real-time schedulers for multi-core processors.

There are several areas we would like to explore in order to improve the current approach. First, we realize that the rounding of frequency under real frequency settings may lead to a waste of processer resources, and in turn result in more energy consumption and even weaker schedulability. Thus we will explore the possibility that takes frequency settings as conditions of energy optimization in order to alleviate the loss on rounding frequency. Secondly, at present we only evaluate our DVS strategies on simulators, upon which many factors could not be simulated precisely compared to the real case. In future, we plan to conduct our DVS strategies on real multi-core processors to produce more accurate performance and energy evaluation results.

% if have a single appendix:
%\appendix[Proof of the Zonklar Equations]
% or
%\appendix  % for no appendix heading
% do not use \section anymore after \appendix, only \section*
% is possibly needed

% use appendices with more than one appendix
% then use \section to start each appendix
% you must declare a \section before using any
% \subsection or using \label (\appendices by itself
% starts a section numbered zero.)
%

%\appendices
%\section{Proof of the First Zonklar Equation}
%Appendix one text goes here.
%
%% you can choose not to have a title for an appendix
%% if you want by leaving the argument blank
%\section{}
%Appendix two text goes here.

%% use section* for acknowledgement
%\ifCLASSOPTIONcompsoc
%  % The Computer Society usually uses the plural form
%  \section*{Acknowledgments}
%\else
%  % regular IEEE prefers the singular form
%  \section*{Acknowledgment}
%\fi
%
%
%The authors would like to thank...
%
%
%% Can use something like this to put references on a page
%% by themselves when using endfloat and the captionsoff option.
%\ifCLASSOPTIONcaptionsoff
%  \newpage
%\fi

% trigger a \newpage just before the given reference
% number - used to balance the columns on the last page
% adjust value as needed - may need to be readjusted if
% the document is modified later
%\IEEEtriggeratref{8}
% The "triggered" command can be changed if desired:
%\IEEEtriggercmd{\enlargethispage{-5in}}

% use section* for acknowledgement
\ifCLASSOPTIONcompsoc
  % The Computer Society usually uses the plural form
  \section*{Acknowledgments}
\else
  % regular IEEE prefers the singular form
  \section*{Acknowledgment}
\fi

This work is supported partly by the National Basic Research Program of China (973) under Grant No. 2009CB320703, the Science Fund for
Creative Research Groups of China under Grant No. 60821003,
the National Natural Science Foundation of China under Grant No.61103026, and the National High Technology Research and Development (863) Program  of China under Grant No 2011AA01A202.

% Can use something like this to put references on a page
% by themselves when using endfloat and the captionsoff option.
\ifCLASSOPTIONcaptionsoff
  \newpage
\fi

% trigger a \newpage just before the given reference
% number - used to balance the columns on the last page
% adjust value as needed - may need to be readjusted if
% the document is modified later
%\IEEEtriggeratref{8}
% The "triggered" command can be changed if desired:
%\IEEEtriggercmd{\enlargethispage{-5in}}

% references section

% can use a bibliography generated by BibTeX as a .bbl file
% BibTeX documentation can be easily obtained at:
% http://www.ctan.org/tex-archive/biblio/bibtex/contrib/doc/
% The IEEEtran BibTeX style support page is at:
% http://www.michaelshell.org/tex/ieeetran/bibtex/
\bibliographystyle{IEEEtran}
% argument is your BibTeX string definitions and bibliography database(s)
\bibliography{IEEEabrv}
%
% <OR> manually copy in the resultant .bbl file
% set second argument of \begin to the number of references
% (used to reserve space for the reference number labels box)
%\begin{thebibliography}{1}

%\bibitem{IEEEhowto:kopka}
%H.~Kopka and P.~W. Daly, \emph{A Guide to \LaTeX}, 3rd~ed.\hskip 1em plus
%  0.5em minus 0.4em\relax Harlow, England: Addison-Wesley, 1999.

%\end{thebibliography}

% biography section
%
% If you have an EPS/PDF photo (graphicx package needed) extra braces are
% needed around the contents of the optional argument to biography to prevent
% the LaTeX parser from getting confused when it sees the complicated
% \includegraphics command within an optional argument. (You could create
% your own custom macro containing the \includegraphics command to make things
% simpler here.)
%\begin{biography}[{\includegraphics[width=1in,height=1.25in,clip,keepaspectratio]{mshell}}]{Michael Shell}
% or if you just want to reserve a space for a photo:

%\begin{IEEEbiography}{Michael Shell}
%Biography text here.
%\end{IEEEbiography}

% if you will not have a photo at all:

% You can push biographies down or up by placing
% a \vfill before or after them. The appropriate
% use of \vfill depends on what kind of text is
% on the last page and whether or not the columns
% are being equalized.

%\vfill

% Can be used to pull up biographies so that the bottom of the last one
% is flush with the other column.
%\enlargethispage{-5in}

\balance

% that's all folks
\end{document}